\documentclass[a4paper,12pt]{article}
\usepackage{cite}
\usepackage[ref]{overcite}
\usepackage{graphicx}
\begin{document}
\baselineskip=0.8truecm
\title{\bf Chaotic ray dynamics in an optical cavity with a beam splitter}

\author{G. Puentes, A. Aiello, J. P. Woerdman \\
 {\em Huygens Laboratory, Leiden University, P.O. Box 9504,
Leiden, The Netherlands}}
\maketitle

\begin{abstract}
We investigate the ray dynamics in an optical cavity when a ray
splitting mechanism is present. The cavity is a conventional
two-mirror stable resonator and the ray splitting is achieved by
inserting an optical beam splitter perpendicular to the cavity
axis. Using Hamiltonian optics, we show that such a simple device
presents a surprisingly rich chaotic ray dynamics.
\end{abstract}

\begin{flushleft}
 \end{flushleft}
\thispagestyle{empty}

\begin{figure}[h]\centerline{\scalebox{0.31}{\includegraphics{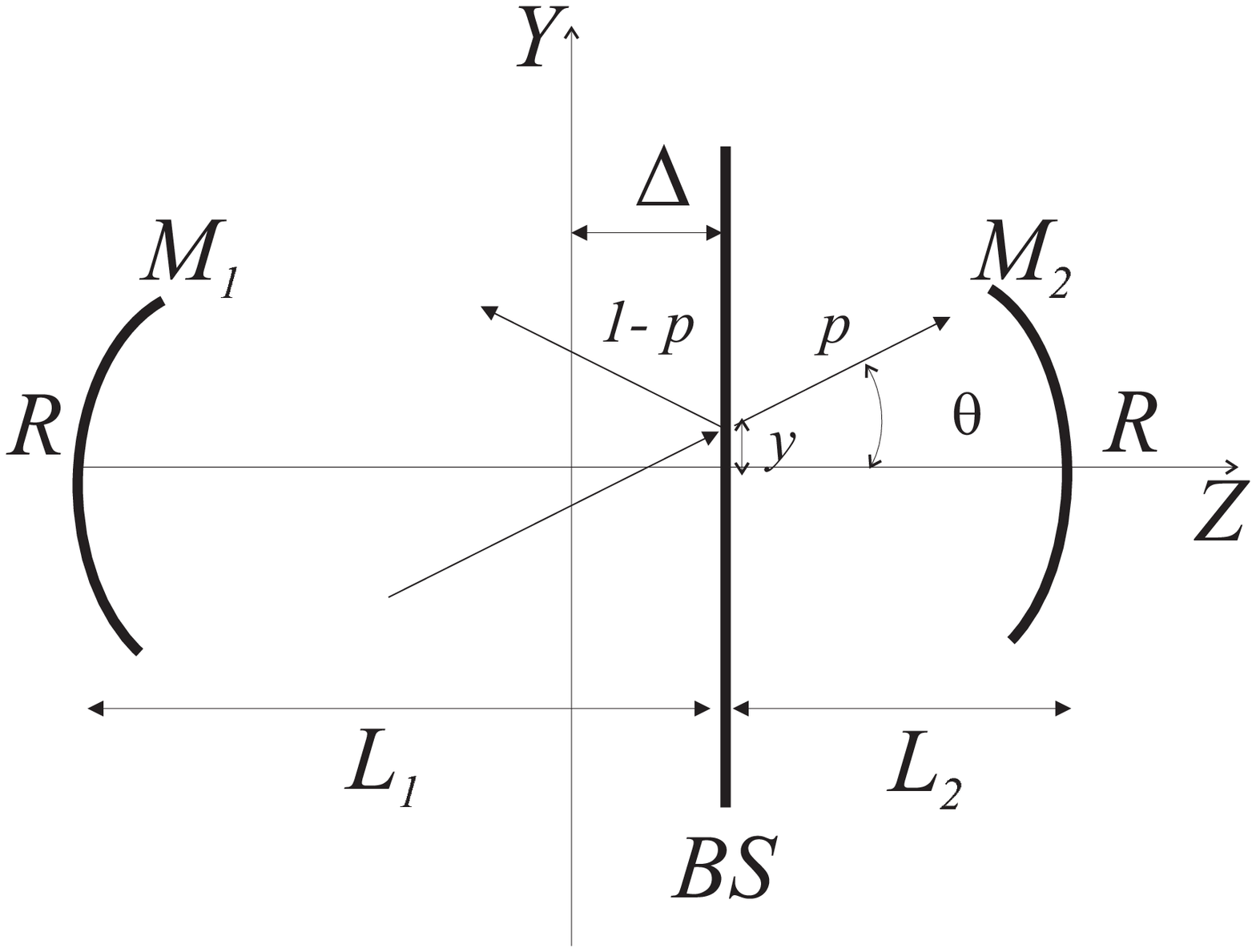}}}
\caption{Schematic diagram of the cavity model; R indicates the
radius of curvature of the mirrors. Two subcavities of length
$L_{1}$ and $L_{2}$ are coupled by a BS. The total cavity is
globally stable for $L=L_1+L_2 < 2 R$. $\Delta=L_1 - L/2$
represents the displacement of the BS with respect to the center
of the cavity. When a ray hits the surface of the BS, which we
choose to coincide with  the reference plane, it can be either
reflected or transmitted with equal probability; for a $50/50$
beam splitter $p=1/2$.} \label{fig:1}
\end{figure}

In this Letter we present a very simple optical cavity whose ray
dynamics is nevertheless fully chaotic. Our starting point is the
fact that a two-mirror optical cavity can be \emph{stable} or
\emph{unstable} depending on its geometrical configuration
\cite{SiegmanBook}. If a light ray is injected inside the cavity
it will remain confined indefinitely when the configuration is
stable but it will escape after a finite number of bounces when
the cavity is unstable. Our interest is in a cavity which has both
aspects of stability and instability (Fig \ref{fig:1}). The cavity
is modelled as a strip resonator \cite{SiegmanBook} made of two
identical concave mirrors of radius of curvature $R$ separated by
a distance $L$, where $L<2R$ so that the cavity is globally
stable. We then introduce a beam splitter (BS) inside the cavity,
oriented perpendicular to the optical axis. In this way the BS
defines two planar-concave subcavities: one on the left and one on
the right with respect to the BS, with length $L_1$ and $L_2$,
respectively. The main idea is that depending on the position of
the BS the left (right) subcavity becomes \emph{unstable} for the
reflected rays when $L_1$ ($L_2$) is bigger than $R$, while the
cavity as a whole remains always \emph{stable} ($L_1+L_2<2R$).

Consideration of this system raises the nontrivial question
whether there will be an "equilibrium" between the number of
trapped rays and escaping rays. The trapped rays are those which
bounce for infinitely long times due to the global stability of
the cavity and the escaping ones are those which stay only for a
finite time. If such equilibrium exists it could eventually lead
to transient chaos since it is known in literature that
instability (positive Lyapunov exponents) and mixing (confinement
inside the system) form the skeleton of chaotic dynamics
\cite{Cvitanovic02}. In this Letter we show that under certain
conditions such equilibrium can be achieved
in our cavity and that chaotic ray dynamics is displayed\\

In our system the BS plays a crucial role. It is modelled as a
stochastic ray splitting element by assuming the reflection and
transmission coefficients as random variables \cite{Couchman92a}.
Within the context of wave optics this model corresponds to the
neglect of all interference phenomena inside the cavity, as
required by the ray (zero-wavelength) limit. The stochasticity is
implemented by using a Monte Carlo method to determine whether the
ray is transmitted or reflected \cite{Couchman92a}. When a ray is
incident on the ray splitting surface of the BS, it is either
transmitted through it, with probability $p$, or reflected with
probability $1-p$, where we assume $p=1/2$ for a $50/50$ beam
splitter as shown in Fig \ref{fig:1}. We then dynamically evolve a
ray and at each reflection we use a random number generator with a
uniform distribution to randomly decide whether
to reflect or transmit the incident ray.\\

\begin{figure}[h]\centerline{\scalebox{0.50}{\includegraphics{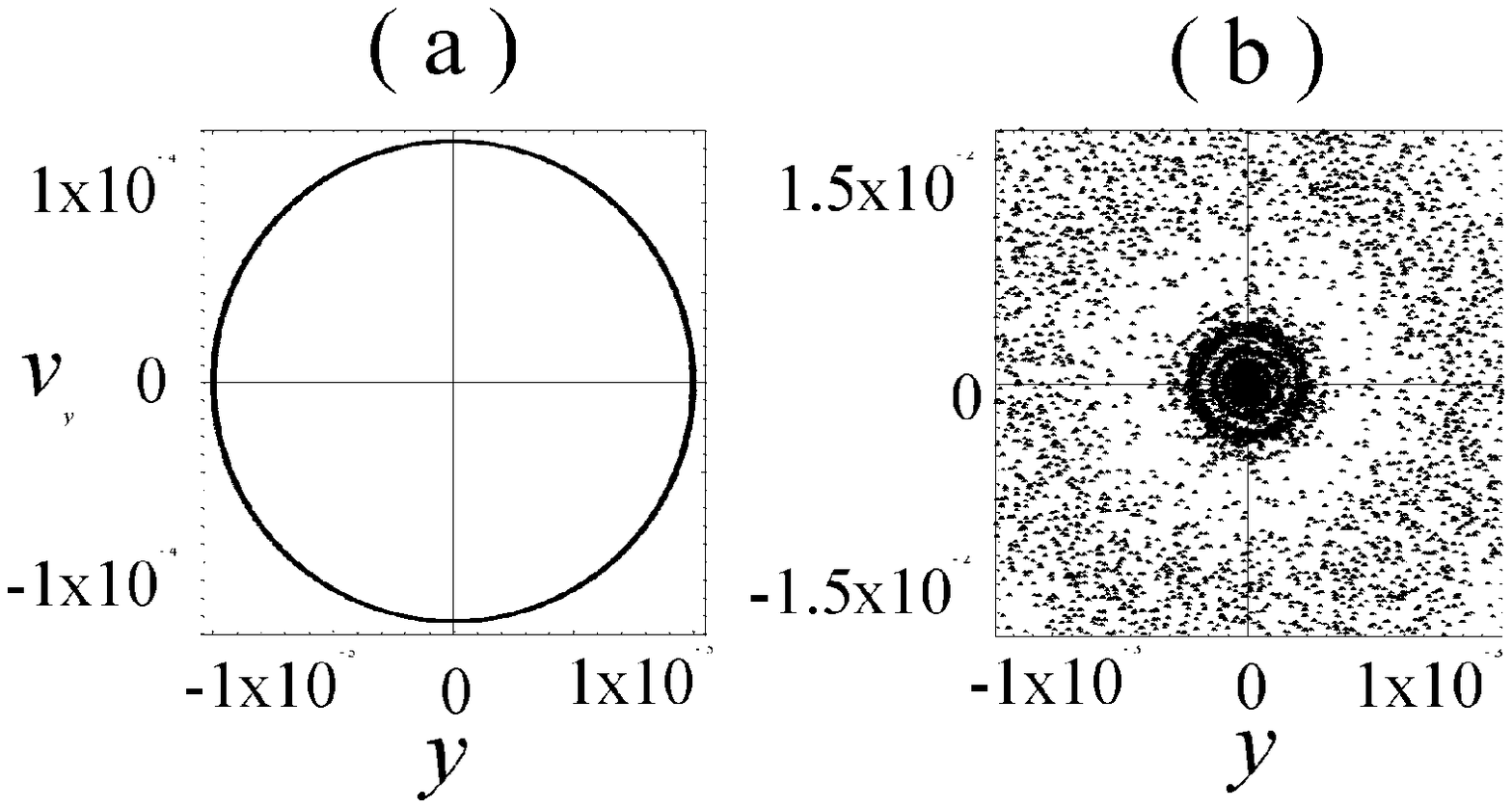}}}
  \caption{SOS for (a) $\Delta=0$: the ray dynamics is stable and thus
confined on a torus in phase space. (b) $\Delta=0.002$m, the
dynamics becomes unstable and the ray escapes after $n=75328$
bounces. Note the ring structure in this plot.}
  \label{fig:2}
  \end{figure}

In the context of Hamiltonian optics, to characterize the
trajectory of a ray we first choose a reference plane
perpendicular to the optical axis $\hat{Z}$, coinciding with the
surface of the BS. The intersection of a ray with this plane is
specified by two parameters: the height $y$ above the optical axis
and the angle $\theta$ between the trajectory and the same axis.
We consider the rays as point particles, as in standard billiard
theory where the propagation of rays correspond to the
trajectories of unit mass point particles moving freely inside the
billiard and reflecting elastically at the boundary. In
particular, we study the evolution of the transversal component of
the momentum of the ray, i.e. $v_{y}=|\vec{v}|\sin(\theta)$ so
that we associate a ray of light with the two-dimensional vector
$\vec{r}=(y,v_y)$. It is important to stress that we use exact
2D-Hamiltonian optics, i.e. we do \emph{not} use the
paraxial approximation. \\
The evolution of a set of rays injected in the cavity with
different initial conditions $(y_0,v_{y_0})$, is obtained by using
a ray tracing algorithm. For each initial condition, the actual
ray trajectory is determined by a random sequence
\{...rrttrrrrt..\} which specifies if the ray is reflected (r) or
transmitted (t) by the BS. When one evolves the whole set of
initial conditions, one can choose between  two possibilities,
either use the {\em same} random sequence for all rays in the set
of initial conditions or use a {\em different} random sequence for
each ray. In this Letter we use the {\em same} random sequence for
all injected rays in order to uncover the
dynamical randomness of the cavity.\\
The three quantities that we have calculated to demonstrate the
chaotic  ray dynamics inside the cavity are the Poincar\'{e}
Surface of Section (SOS), the exit basin diagrams and the escape
time function \cite{OttBook}. In all calculations we have assumed
$L_1+L_2=0.16$m and the radius of curvature of the mirrors
$R=0.15$m; the diameter $d$ of the two mirrors was $d=0.05$m. In
addition, the displacement $\Delta$ of the BS  with respect to the
center of the cavity was chosen as $0.02$m (unless specified
otherwise),
and the time was measured in number of bounces $n$.\\
In Fig. \ref{fig:2}, the successive intersections of a ray with
initial transverse coordinates $y_0=1\times10^{-5}$m, $v_{y_0}=0$
are represented by the  black points in the SOSs. For $\Delta=0$
the cavity configuration is symmetric and the dynamics is
completely regular (Fig.\ref{fig:2}(a)); the on-axis trajectory
represents an elliptic fixed point and nearby stable trajectories
lie on continuous tori in phase space. In Fig.\ref{fig:2}(b), the
BS is slightly displaced from the center ($\Delta=0.02$m), the
same initial trajectory becomes unstable  and spreads over a
finite region of the phase space before escaping after a large
number of bounces ($n=75328$). In view of the ring structure of
Fig.\ref{fig:2}(b) we may qualify the motion as azimuthally
ergodic. The fact that the ray-splitting mechanism introduced by
the BS produces ergodicity is a well known result for a closed
billard \cite{Couchman92a}. We find here an analogue phenomenon,
with the difference that in our case the trajectory does not
explore uniformly but only azimuthally the available phase space,
as an apparent consequence of the openness
of the system.\\

\begin{figure}[h]\centerline{\scalebox{0.37}{\includegraphics{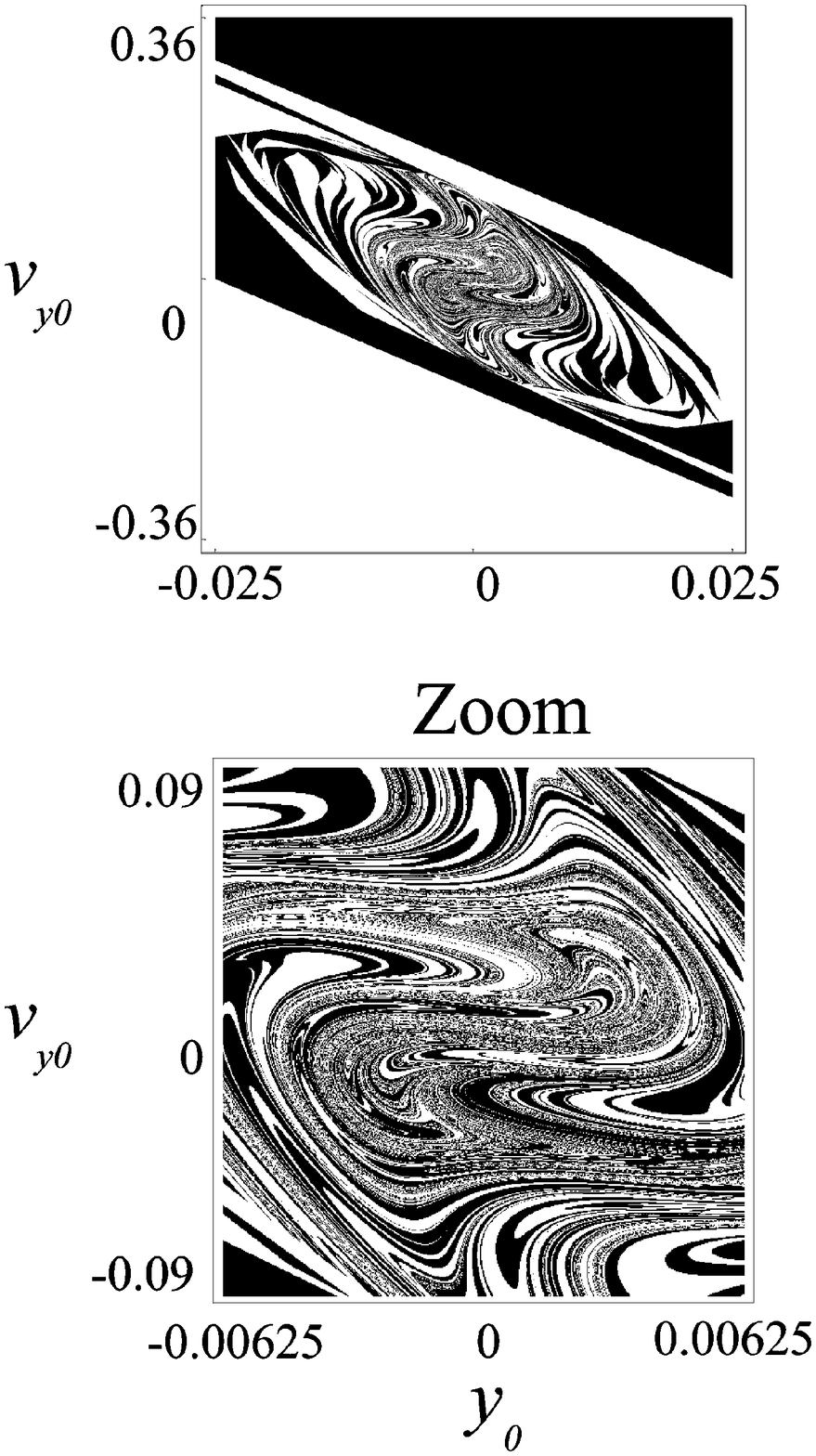}}}
  \caption{Exit basin for $\Delta=0.02$m. The fractal boundaries are a
typical feature of chaotic scattering systems.}
  \label{fig:3}
  \end{figure}

It is well known that chaotic hamiltonian systems with more than
one exit channel exhibit irregular escape dynamics which can be
displayed, e.g., by plotting the exit basin diagram
\cite{Bleher88a}. In our system, this diagram was constructed by
defining a fine grid ($2200\times2200$) of initial conditions
$(y_0,v_{y_0})$. Each ray is followed until it escapes from the
cavity. When it escapes from above ($v_{y}>0$) we plot a black dot
in the corresponding initial condition, whereas when it escapes
from below ($v_{y}<0$) we plot a white dot. This is shown in
Fig.\ref{fig:3}, the uniform regions  in the exit basin diagram
correspond to rays which display a regular dynamics, whereas the
dusty regions correspond to portions of phase space where there is
sensitivity to initial conditions, since two initially nearby
points can escape from opposite exits. Moreover, in
Fig.\ref{fig:3} one can see how the boundary between black and
white regions becomes less and less smooth as one approaches the
center of these diagrams. It is known that this boundary is
actually a fractal set \cite{Ree02a} whose convoluted appearance
is a typical feature of chaotic scattering systems \cite{ottnat}.\\

\begin{figure}[h]\centerline{\scalebox{0.35}{\includegraphics{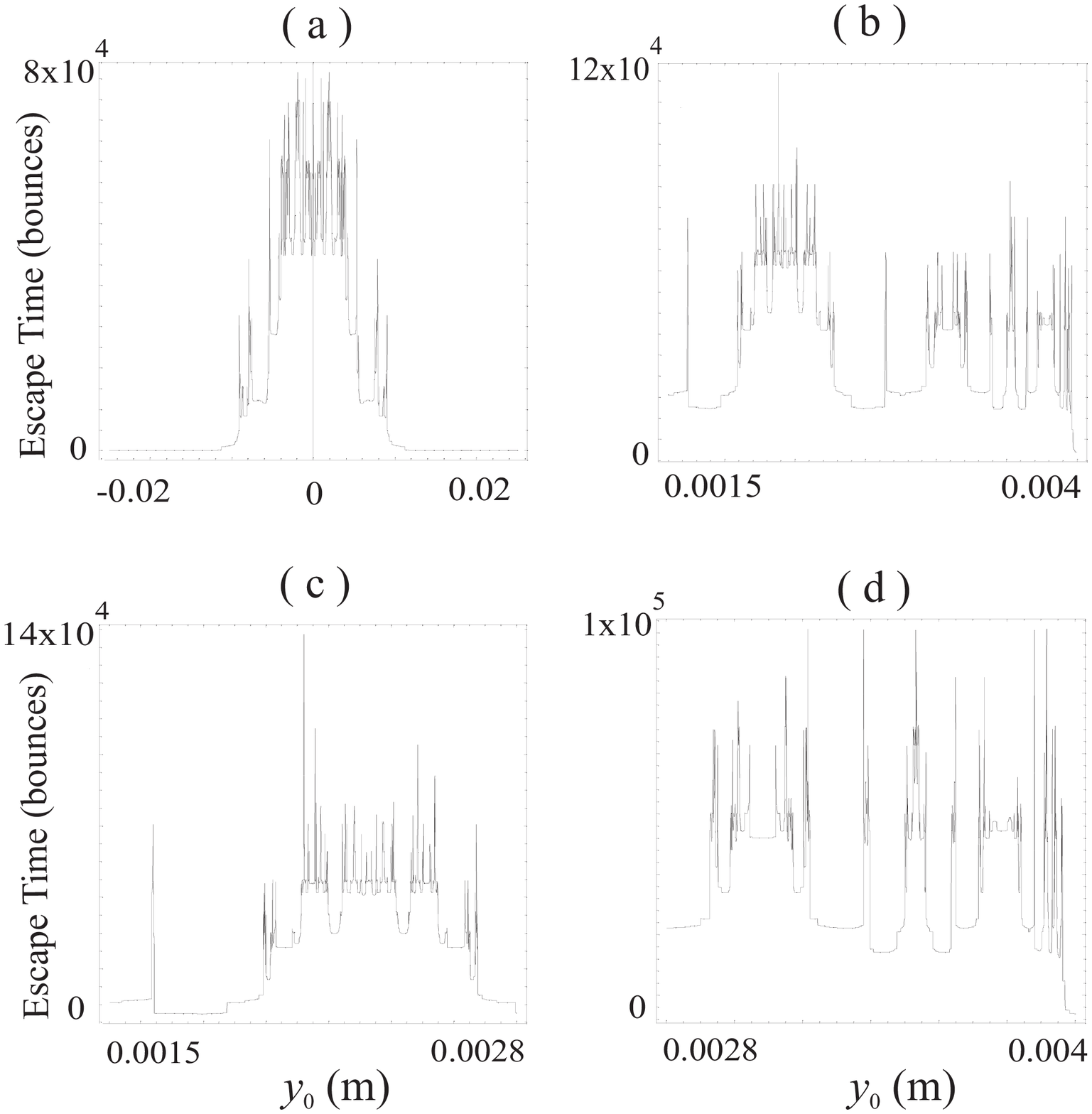}}}
  \caption{(a) Escape time as a function of the initial condition
$y_0$. (b) Blow up of a small interval along the horizontal axis
in (a). (c) and (d) Blow ups of consecutive intervals along the
set of impact parameters $y_0$ shown in (b).}
  \label{fig:4}
  \end{figure}

Besides sensitivity to initial conditions, another fundamental
ingredient of chaotic dynamics is the presence of infinitely long
living orbits which are responsible for the mixing properties of
the system. This set of orbits is usually  called repeller
\cite{GaspardBook}, and is fundamental to generate a truly chaotic
scattering system. To verify the existence of this set we have
calculated  the escape time  or time delay function
\cite{Bleher90D} for a one-dimensional set of initial conditions
specified by the initial position $y_{0}$ (impact parameter) taken
on the mirror $M_1$ and the initial velocity $v_{y_0}=0$. The
escape time was calculated in the standard way, as the time
(in number of bounces $n$) it takes a ray to escape from the cavity.\\
Fig.\ref{fig:4}(a) shows the escape time function. The
singularities of this function are a clear signature of the
existence of long living orbits and the presence of peaks followed
by flat regions are a signature of the exponential sensitivity to
initial conditions. In order to verify the presence of an
\textit{infinite} set of long living orbits, we have zoomed in on
the set of impact parameters $y_{0}$ in three different intervals
(Fig. \ref{fig:4}(b), (c) and (d)). Each zoom reveals the
existence of new infinitely long living orbits. Infinite delay
times correspond to orbits that are asymptotically close to an
unstable periodic orbit. If we would continue to increase the
resolution we would find more and more infinitely trapped orbits.
The repeated existence of singular points is a signature of the
mixing mechanism of the
system due to the global stability of the cavity.\\

In conclusion, we have demonstrated that our simple optical system
displays chaotic ray dynamics. It is important to stress that a
key component for the development of  chaos  is the inclusion of
non-paraxial rays  which add the mixing properties to the system
\cite{aiello}. In fact, it has been previously shown that paraxial
ray dynamics can be unstable but not chaotic, in systems with
stochastic perturbations \cite{Longhi02a,puentesPRE}. In our case,
it is the stochastic ray splitting mechanism induced by the BS
that destroys the regular motion of rays in the globally stable
(but non-paraxial) cavity, as shown by the SOSs. Moreover, by
calculating the exit basin diagrams we have found that they show
fractal boundaries, which is a typical feature of chaotic ray
dynamics \cite{Bleher88a}. Finally, through the singularities in
the escape time function, we have verified the presence of
infinitely long living orbits, which in turns reveal the  mixing
mechanism of our optical cavity. An experimental confirmation of
the fractal properties of the exit basin can be performed, e.g.,
in the way suggested in \cite{ottnat}, by injecting a narrow laser
beam into the cavity either in a regular or in a dusty region of
phase space. In the former case one expects the beam to leave the
cavity either from above or below, while in the latter case both
exits should appear illuminated. This proposed experiment is fully
within the context of geometrical optics
(interference plays no role) so that our stochastic model of the BS is adequate.\\

This project is part of the program of FOM and is also supported
by the EU under the IST-ATESIT contract. We thank S. Oemrawsingh
for useful contributions to software development.\\

\newpage

\section*{List of Figure Captions}

Fig. 1.Schematic diagram of the cavity model; R indicates the
radius of curvature of the mirrors. Two subcavities of length
$L_{1}$ and $L_{2}$ are coupled by a BS. The total cavity is
globally stable for $L=L_1+L_2 < 2 R$. $\Delta=L_1 - L/2$
represents the displacement of the BS with respect to the center
of the cavity. When a ray hits the surface of the BS, which we
choose to coincide with  the reference plane, it can be either
reflected or transmitted with equal probability; for a $50/50$
beam splitter $p=1/2$.

Fig. 2.SOS for (a) $\Delta=0$: the ray dynamics is stable and thus
confined on a torus in phase space. (b) $\Delta=0.002$m, the
dynamics becomes unstable and the ray escapes after $n=75328$
bounces. Note the ring structure in this plot.

Fig. 3.Exit basin for $\Delta=0.02$m. The fractal boundaries are a
typical feature of chaotic scattering systems.

Fig. 4.(a) Escape time as a function of the initial condition
$y_0$. (b) Blow up of a small interval along the horizontal axis
in (a). (c) and (d) Blow ups of consecutive intervals along the
set of impact parameters $y_0$ shown in (b).

\newpage

\end{document}